\documentclass{article}

\PassOptionsToPackage{numbers, compress}{natbib}
\usepackage[preprint]{nips_2018}

\usepackage[utf8]{inputenc} % allow utf-8 input
\usepackage[T1]{fontenc}    % use 8-bit T1 fonts
\usepackage{hyperref}       % hyperlinks
\usepackage{url}            % simple URL typesetting
\usepackage{booktabs}       % professional-quality tables
\usepackage{amsfonts}       % blackboard math symbols
\usepackage{nicefrac}       % compact symbols for 1/2, etc.
\usepackage{microtype}      % microtypography
\usepackage[cmex10]{amsmath}
\usepackage{amssymb}
\usepackage{graphicx}

\title{TinySpeech: Attention Condensers for Deep Speech Recognition Neural Networks on Edge Devices}

\author{
  Alexander Wong$^{1,2,3,*}$, Mahmoud Famouri$^{3}$, Maya Pavlova$^{1,3}$, and Siddharth Surana$^{3}$\\
  $^{1}$ Vision and Image Processing Research Group, University of Waterloo, Waterloo, ON, Canada\\
  $^{2}$ Waterloo Artificial Intelligence Institute, University of Waterloo, Waterloo, ON, Canada\\
  $^{3}$ DarwinAI Corp., Waterloo, ON, Canada \\
  \texttt{$^{*}$ a28wong@uwaterloo.ca} \\
}

\begin{document}

\maketitle

\begin{abstract}
  Advances in deep learning have led to state-of-the-art performance across a multitude of speech recognition tasks.  Nevertheless, the widespread deployment of deep neural networks for on-device speech recognition remains a challenge, particularly in edge scenarios where the memory and computing resources are highly constrained (e.g., low-power embedded devices) or where the memory and computing budget dedicated to speech recognition is low (e.g., mobile devices performing numerous tasks besides speech recognition).  In this study, we introduce the concept of \textbf{attention condensers} for building low-footprint, highly-efficient deep neural networks for on-device speech recognition on the edge.  An attention condenser is a self-attention mechanism that learns and produces a condensed embedding characterizing joint local and cross-channel activation relationships, and performs selective attention accordingly.  Unlike self-attention mechanisms designed for deep convolutional neural networks that depend heavily on existing convolution modules, attention condensers act as self-contained, stand-alone modules and facilitate for efficient deep neural networks with much sparser use of larger stand-alone convolution modules and more frequent use of attention condensers.  To illustrate its efficacy, we introduce \textbf{TinySpeech}, low-precision deep neural networks comprising largely of attention condensers tailored for on-device speech recognition using a machine-driven design exploration strategy, with one tailored specifically with microcontroller operation constraints.  Experimental results on the Google Speech Commands benchmark dataset for limited-vocabulary speech recognition showed that TinySpeech networks achieved significantly lower architectural complexity (as much as $507\times$ fewer parameters), lower computational complexity (as much as $48\times$ fewer multiply-add operations), and lower storage requirements (as much as $2028\times$ lower weight memory requirements) when compared to previous work.  These results not only demonstrate the efficacy of attention condensers for building highly efficient deep neural networks for on-device speech recognition, but also illuminate its potential for accelerating deep learning on the edge and empowering a wide range of TinyML applications.

\end{abstract}

\section{Introduction}
%\vspace{-0.15in}
\label{Introduction}

Advances in deep learning~\cite{lecun2015deep} have led to significant improvements in a plethora of complex tasks, ranging from visual perception~\cite{krizhevsky2012imagenet,he2015deep} to natural language processing~\cite{vaswani2017attention,devlin2018bert} to drug discovery~\cite{alperstein2019smiles,wallach2015atomnet}.  A particular area of interest where deep learning has shown exceptional performance beyond other machine learning strategies has been speech recognition~\cite{hannun2014deep,xiong2016achieving,Warden2018,Sainath2015,Tang2017,Tang2017_Honk,myer2018efficient,anderson2020performanceoriented}, where it has demonstrated state-of-the-art performance across a multitude of speech recognition tasks ranging from conversational speech recognition to limited-vocabulary speech recognition.  Deep learning for speech recognition has been so successful that it is now widely used in real-world applications ranging from voice assistants (e.g., Amazon Alexa, Microsoft Cortana, Google Assistant, Apple Siri) to real-time closed captioning (e.g., Youtube Closed Captions, Microsoft Teams Live Captions).

Despite the successes of deep learning in speech recognition from an accuracy perspective, widespread deployment of deep neural networks for on-device speech recognition remains a challenge, particularly in edge scenarios where the memory and computing resources are highly constrained (e.g., low-power embedded devices) or where the memory and computing budget dedicated to speech recognition is very low (e.g., mobile devices performing numerous tasks besides speech recognition).  Taking a step forward in the area of on-device speech recognition, there has been significant research interest in recent years on limited-vocabulary speech recognition~\cite{Warden2018}, where the objective is to recognize words from verbal utterances within a limited vocabulary of words. The ability to perform real-time, on-device limited-vocabulary speech recognition can enable widespread proliferation of voice interfaces on low-cost, low-power edge devices without the need for cloud computing, which is critical in situations where privacy and security is paramount and in situations where bandwidth and connectivity is limited.

Given the significant interest in on-device speech recognition on the edge, there has been a much greater focus in recent years on the design of low-footprint, highly-efficient deep neural networks for on-device limited-vocabulary speech recognition~\cite{Warden2018,Sainath2015,Tang2017,Tang2017_Honk,myer2018efficient,anderson2020performanceoriented,blouw2020hardware}.  These low-footprint deep neural network designs have centered around leveraging well-known deep convolutional neural network design patterns (e.g.,~\cite{he2015deep}) to construct low-complexity architectures, and have been shown to achieve strong recognition accuracy while maintaining low architectural and computational complexity.  However, there are complexity barriers that limit how efficient deep neural networks based on existing deep convolutional neural network design patterns can achieve, and as such exploring alternative design patterns that can achieve even greater efficiency is highly desired.

Motivated to push the complexity barrier even lower than possible with existing deep convolutional neural network design patterns, in this study we introduce the concept of \textbf{attention condensers} for building highly-efficient yet high-performance deep neural networks for speech recognition on edge devices.  More specifically, an attention condenser is a self-attention mechanism that learns and produces a condensed embedding characterizing joint local cross-channel activation relationships, and performs selective attention accordingly with a greater emphasis on activations in close proximity of strong activations.  Unlike self-attention mechanisms designed for deep convolutional neural networks that depend heavily on existing convolution modules, attention condensers act as self-contained, stand-alone modules and facilitate for efficient deep neural networks with much sparser use of stand-alone convolution modules and more frequent use of attention condensers.  Incorporating the proposed attention condensers within a machine-driven design exploration strategy, we introduce low-precision deep neural networks comprising largely of attention condensers, nicknamed \textbf{TinySpeech}, tailored specifically for limited-vocabulary speech recognition.

The paper is organized as follows. In Section~\ref{Methods}, a detailed description of the underlying theory behind attention condensers and the proposed TinySpeech deep neural network architectures are presented.  In Section~\ref{Results}, experimental results are presented where we evaluated the efficacy and efficiency of the proposed TinySpeech networks experimentally on the Google Speech Commands benchmark dataset for limited-vocabulary speech recognition.  Conclusions are drawn and future work discussed in Section~\ref{Conclusions}.  The broader impact of this work is discussed in Section~\ref{BroaderImpact}. 

\section{Methods}
%\vspace{-0.15in}
\label{Methods}
\begin{figure}
%\vspace{- 0.6 cm}
\centering
	\includegraphics[width = \linewidth]{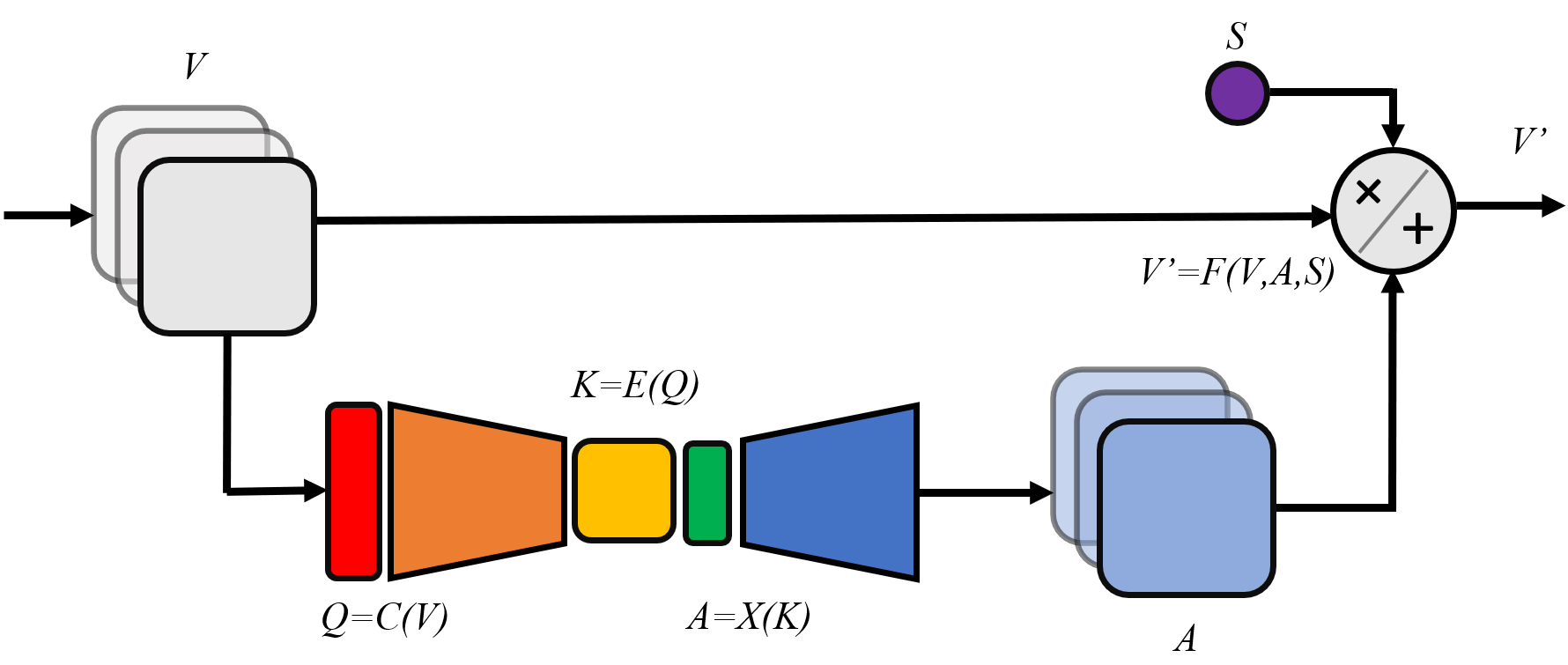}
	\caption{An attention condenser is a self-attention mechanism consisting of a condensation layer $C(V)$, an embedding structure $E(Q)$, an expansion layer $X(K)$, and a selective attention mechanism $F(V,A,S)$.  The condensation layer $C(V)$ condenses the input activations $V$ for reduced dimensionality to $Q$ in a way that emphasizes activations in close proximity of strong activations.  An embedding structure $E(Q)$ then learns and produces a condensed embedding $K$ from $Q$ that characterizes joint local and cross-channel activation relationships.  An expansion layer $X(K)$ projects the condensed embedding $K$ for increased dimensionality to produce self-attention values $A$ to be used for selective attention $F(V,A,S)$.  The output $V'$ are a product of the input activations $V$, self-attention values $A$, and scale $S$ via selective attention $F(V,A,S)$.}	
	\label{fig:condenser}
%\vspace{-0.6 cm}
\end{figure}

In this study, we leveraged two complementary strategies to build \textbf{TinySpeech}, low-footprint, low-precision deep neural networks tailored specifically for limited-vocabulary speech recognition.  First, we introduce the concept of \textbf{attention condensers}, a new self-attention mechanism designed for selective attention based on joint local and cross-channel activation relationships captured via condensed embeddings.  Second, machine-driven design exploration incorporating this new attention condenser is leveraged to automatically determine the macroarchitecture and microarchitecture designs of the final TinySpeech networks for optimal balance between speech recognition accuracy and network efficiency.  Details of these two strategies, along with details about the TinySpeech network architectures, are described below.

\subsection{Attention Condensers}
%\vspace{-0.1in}

To push the complexity barrier even lower than possible with existing deep convolutional neural network design patterns, the first strategy we took in building TinySpeech is the introduction of a new self-attention mechanism called attention condensers.  The use of self-attention mechanisms in deep learning have grown significantly in popularity due to their effectiveness in recent years~\cite{vaswani2017attention,bahdanau2014neural,velickovic2018graph,pmlr-v97-zhang19d,hu2017squeezeandexcitation,woo2018cbam}, particularly in the realm of natural language processing with the advent of Transformers~\cite{vaswani2017attention}, where they demonstrated superior performance by solely using stand-alone self-attention mechanisms without leveraging recurrence or convolutions in the network architecture.  Self-attention mechanisms have also been investigated and explored to augment deep convolutional neural network architectures with great effect~\cite{hu2017squeezeandexcitation,woo2018cbam}.  Existing self-attention mechanisms for deep convolutional neural network architectures in literature have focused on the decoupling of attention into channel-wise attention~\cite{hu2017squeezeandexcitation} and local attention~\cite{woo2018cbam}, where channel-wise attention mechanisms project input activations along the non-channel dimensions and model cross-channel activation relationships while local attention mechanisms project input activations along the channel dimension and model local activation relationships.  Furthermore, existing self-attention mechanisms for deep convolutional neural network architectures are designed to depend on existing convolution modules within the network architecture. As such, while existing self-attention mechanisms are designed to augment network architectures to improve accuracy at the expense of some complexity, they are not designed to be stand-alone mechanisms for facilitating improved network efficiency.

Motivated to create a self-contained, stand-alone self-attention mechanism that facilitate for sparser use of larger stand-alone convolution modules to reduce overall network complexity, we designed the proposed attention condensers in a way that jointly models both local and cross-channel activation relationships within a unified embedding.  To greatly reduce the computational complexity of such a joint modeling, we condense the input activations to a reduced dimension to strike a balance between modeling capability and computational efficiency.  An overview of the proposed attention condenser is shown in Figure~\ref{fig:condenser}.  More specifically, an attention condenser is a self-attention mechanism consisting of a condensation layer $C(V)$, an embedding structure $E(Q)$, an expansion layer $X(K)$, and a selective attention mechanism $F(V,A,S)$.  The condensation layer (i.e., $Q=C(V)$) condenses the input activations $V$ for reduced dimensionality to $Q$ in a way that emphasizes activations in close proximity to strong activations.  An embedding structure (i.e., $K=E(Q)$) then learns and produces a condensed embedding $K$ from $Q$ that characterizes joint local and cross-channel activation relationships.  An expansion layer (i.e., $A=X(K)$) projects the condensed embedding $K$ for increased dimensionality to produce self-attention values $A$ to be used for selective attention (i.e., $V'=F(V,A,S)$).  The output $V'$ are a product of the input activations $V$, self-attention values $A$, and scale $S$ via selective attention (i.e., $V'=F(V,A,S)$).  The introduction of scale $S$ to the selective attention mechanism enables greater flexibility for the degree of selective attention to be determined.  More specifically, the degree of selective attention increases as $S$ decreases such that as $S\rightarrow 0$, $V'\rightarrow A$.  Overall, by jointly modeling local and cross-channel activation relationships within a unified condensed embedding, attention condensers can act as self-contained, stand-alone modules and facilitate for efficient deep neural networks with much sparser use of larger stand-alone convolution modules and more frequent use of attention condensers.

\subsection{Machine-driven Design Exploration}
%\vspace{-0.1in}
Given the proposed attention condenser, we now leverage a machine-driven design exploration strategy that incorporates this new attention condenser to automatically determine the macroarchitecture and microarchitecture designs of the final TinySpeech network architectures to tailor it specifically for the purpose of on-device limited-vocabulary speech recognition with an optimal balance between speech recognition accuracy and network efficiency.

Motivated by past literature in the area of deep convolutional neural network architectures for limited-vocabulary speech recognition, we leverage mel-frequency cepstrum coefficient (MFCC) representations, derived from the input audio signal, as the input to TinySpeech.  More specifically, as per~\cite{Tang2017_Honk}, we leverage a two-dimensional stack of MFCC representations with a 30ms window and a 10ms time shift across a one-second audio sample that has been band-pass filtered with cutoff from 20Hz to 4kHz for reducing noise.  This two-dimensional stack of MFCC representations enable the capturing of time-frequency characteristics of the input signal.  For learning a condensed embedding that characterizes joint local time-frequency and cross-channel activation relationships in an efficient yet effective manner, we leveraged max pooling, a lightweight two-layer neural network (grouped then pointwise convolution), and unpooling for the condensation layer $C(V)$, the embedding structure $E(Q)$, and the expansion layer $X(K)$, respectively, within an attention condenser.

To perform machine-driven design exploration for automatically determining the macroarchitecture and microarchitecture design of the final TinySpeech network architecture, we incorporated the new attention condenser design pattern into generative synthesis~\cite{Wong2018}, a highly flexible generative approach to creating highly tailored deep neural network architectures based on operational requirements and constraints.  Amongst the machine-driven design exploration strategies in literature~\cite{anderson2020performanceoriented,zoph2018learning,cai2019proxylessnas,tan2019mnasnet}, the generative synthesis approach is well-suited for performing design exploration in this study given that it facilitates very fine-grained macroarchitecture and microarchitecture exploration tailored around both task at hand and operational requirements (e.g., memory footprint, computational efficiency, accuracy, etc.) and the ultimate goal of TinySpeech is to produce a highly-efficient deep neural network for on-device speech recognition in computational and memory constrained scenarios such as on low-cost, low-power edge devices.   For the sake of brevity (a detailed description of generative synthesis can be found in~\cite{Wong2018}), the concept of generative synthesis revolves around solving a constrained optimization problem, where we wish to find a generator $\mathcal{G}$ whose generated deep neural network architectures $\left\{N_s|s \in S\right\}$, with $S$ denoting a set of seeds, maximize a universal performance function $\mathcal{U}$ (e.g.,~\cite{Wong2018_Netscore}), constrained by an indicator function $1_r(\cdot)$ that encapsulates operational requirements,
%\vspace{-0.05in}
\begin{equation}
\mathcal{G}  = \max_{\mathcal{G}}~\mathcal{U}(\mathcal{G}(s))~~\textrm{subject~to}~~1_r(\mathcal{G}(s))=1,~~\forall s \in S.
\label{optimization}
%\vspace{-0.17in}
\end{equation}
To solve this constrained optimization problem in a tractable manner given the enormous space of possible solutions, generative synthesis finds an approximate solution through an iterative process, where an initial solution is defined based on a prototype $\varphi$, $\mathcal{U}$, and $1_r(\cdot)$, with a number of successive generators being constructed.  Given that the goal is to create a highly efficient deep neural network architecture tailored for limited-vocabulary speech recognition in edge scenarios, we define $1_r(\cdot)$ as follows.  In the first experiment, the indicator function $1_r(\cdot)$ is defined such that: i) the validation accuracy is greater than or equal to 90\% on the Google Speech Commands dataset~\cite{Warden2017}, a benchmark dataset designed specifically for limited-vocabulary speech recognition, ii) number of parameters < 15k, and iii) 8-bit weight precision.  A validation accuracy constraint of 90\% validation accuracy was chosen to make TinySpeech comparable in accuracy to a deep neural network proposed in~\cite{Sainath2015} for on-device limited-vocabulary speech recognition that is commonly used as a baseline reference.  A parameter constraint of less than 15k was chosen to ensure that the resulting deep neural network has a small memory footprint given strict memory constraints in edge scenarios.  An 8-bit weight precision constraint was chosen to account for the memory constraints of edge scenarios.  For an additional experiment, we incorporated an additional microcontroller constraint such that the deep neural architecture must only be comprised of operations from the very limited set of operations supported by TensorFlow Lite for Microcontrollers.  Such a microcontroller operation constraint was imposed for this additional experiment to enable us to explore the macroarchitecture and microarchitecture designs under a much more constrained scenario specifically tailored for the purpose of microcontroller deployment, which is a very important deployment scenario for on-device limited-vocabulary speech recognition.

Given that a number of successive generators are being constructed during the generative synthesis process, for the first experiment we take three of the constructed generators at different stages to automatically generate three highly efficient deep speech recognition networks (TinySpeech-X, TinySpeech-Y, and TinySpeech-Z) with different performance-efficiency tradeoffs.  Furthermore, for the additional experiment, one of the constructed generators was leveraged to automatically generate a fourth deep speech recognition network (TinySpeech-M).

\begin{figure}
%\vspace{- 0.6 cm}
\centering
	\includegraphics[width = 0.7\linewidth]{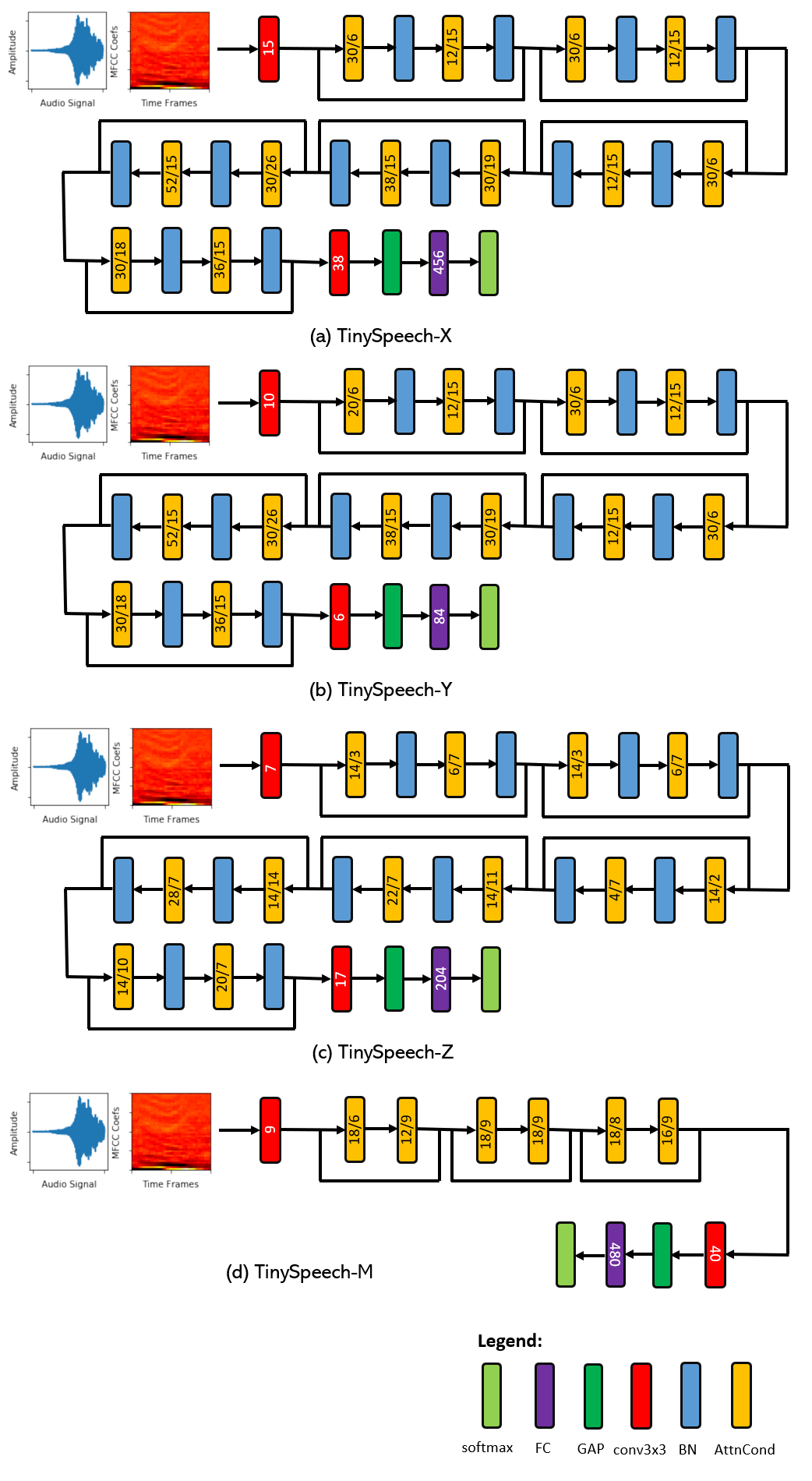}
	\caption{TinySpeech architectures for limited-vocabulary speech recognition.  The number in each convolution module represents the number of channels, and the numbers in each attention condenser represents the number of channels for the first layer and second layer of the embedding structure, respectively, and the number in each fully-connected layer represents the number of synapses.
The TinySpeech network architectures take a MFCC representation of an audio signal as input, and comprise of a convolutional layer, a stack of attention condensers for consecutive selective attention, a second convolutional layer, an global average pooling layer, a fully-connected layer, and finally a softmax layer.  The TinySpeech architectures exhibit high architectural diversity (e.g., heterogeneous mix of  modules with very different architecture designs), as well as a very heavy use of attention condensers and sparse use of stand-alone convolution modules.  These characteristics result in highly efficient deep neural network architectures tailored for edge scenarios.  Furthermore, the TinySpeech-M network architecture, due to the additional microcontroller operation constraints, is much shallower than the other TinySpeech architectures and does not leverage batch normalization.}	
	\label{fig:tinyspeechnet}
\vspace{-0.21 cm}
\end{figure}

For the prototype $\varphi$, we define a residual design prototype whose input is a two-dimensional stack of MFCC representations, and the last layers being an average pooling layer, followed by a fully-connected layer, and a softmax layer indicating the identified word from the verbal utterance.  A residual design prototype was leveraged here given that residual deep convolution neural network architectures have been shown to achieve state-of-the-art speech recognition performance for limited-vocabulary speech recognition~\cite{Tang2017}, and thus well-suited for defining the initial solution.  The use of attention condensers are not defined in the prototype $\varphi$ in order to give full flexibility to the machine-driven design exploration process to automatically determine how and where to best leverage them within a deep neural network architecture for satisfying operational requirements.  As such, the macroarchitecture and microarchitecture design of the final TinySpeech network architecture is left for the machine-driven design exploration process to automatically determine by solving this constrained optimization problem using generative synthesis.

\subsection{Final Architecture Design}
%\vspace{-0.1in}
The final architecture design of the four TinySpeech networks produced by the generative synthesis machine-driven design exploration strategy is shown in Figure~\ref{fig:tinyspeechnet}.  The most interesting and obvious observation about the TinySpeech network architectures is the sparse appearance of stand-alone convolution modules.  In fact, there are a total of two stand-alone convolution modules, with one being the input layer of TinySpeech and the other being the layer feeding into the global average pooling layer.  Instead, the TinySpeech network architectures are largely comprised of consecutive attention condensers, which in this case learns and produces a stacked condensed embedding that characterizes joint local time-frequency and cross-channel activation relationships for the purpose of selective concentration.  The heavy use of attention condensers and sparse use of stand-alone convolution modules result in significantly lower computational complexity in the resulting network architectures.  Another interesting observation about the TinySpeech network architectures is the high architectural diversity. For example, it can be seen from Figure~\ref{fig:tinyspeechnet} that the network architectures consists of a heterogeneous mix of stand-alone convolution modules, attention condensers, and fully-connected modules with very different microarchitecture designs, and the microarchitecture designs are quite different between the TinySpeech networks.  This level of architectural diversity can only be accomplished by leveraging a fine-grained machine-driven design exploration such as generative synthesis.  Furthermore, it can be observed that the TinySpeech architectures have very lightweight architectures with very low architecture complexities, and is a result of the strict parameter constraint imposed during the machine-driven design exploration process.  These observations about the characteristics of the TinySpeech network architectures reveal highly efficient deep neural network architectures tailored for edge scenarios.  Finally, it can be also observed that the TinySpeech-M network architecture, due to the additional microcontroller operation constraints, is much shallower than TinySpeech-X, TinySpeech-Y, and TinySpeech-Z architectures and do not leverage batch normalization (since it is not supported by TensorFlow Lite for Microcontrollers).  This illustrates that the types of constraints imposed on the machine-driven design exploration process can have a significant influence over the architecture of the produced deep neural networks. 

\section{Results and Discussion}
%\vspace{-0.15in}
\label{Results}
\begin{table}[h]
	\centering
	\caption{Test accuracy, number of parameters, and number of multiply-add operations of TinySpeech networks in comparison to five efficient deep speech recognition networks (trad-fpool13~\cite{Sainath2015}, tpool2~\cite{Sainath2015}, res15-narrow~\cite{Tang2017}, TDNN~\cite{myer2018efficient}, PONAS-kws2~\cite{anderson2020performanceoriented}).  Best results are in \textbf{bold}. Results for TinySpeech networks based on 8-bit low precision weights, while results for other tested networks based on 32-bit full precision weights}
	\begin{tabular}{p{3cm}cc|cc}
		\hline
		\textbf{Model}& \textbf{Test Accuracy} & \textbf{Params} &	\textbf{Mult-Adds} \\
		\hline %\hline
		trad-fpool13~\cite{Sainath2015}	&	$90.5\%$	&	1370K	&	125M \\
		tpool2~\cite{Sainath2015}		&	$91.7\%$	&	1090K	&	103M \\
		TDNN~\cite{myer2018efficient} 	&	$94.2\%$ 	&	251K 	&	25.1M \\
		res15-narrow~\cite{Tang2017} 	&	$94.0\%$ 	&	42.6K 	&	160M \\
		PONAS-kws2~\cite{anderson2020performanceoriented} 	&	94.3\% 	&	131K 	&	168M \\
		\hline
		TinySpeech-X		& 	\textbf{94.6\%}	&	10.8K	&	10.9M \\
		TinySpeech-Y		& 	93.6\%	&	6.1K	&	6.5M \\
		TinySpeech-Z		& 	92.4\%	&	\textbf{2.7K}	&	\textbf{2.6M} \\
		TinySpeech-M		& 	91.9\%	&	4.7K	&	4.4M \\
		\hline
	\end{tabular}\\
	\label{tab_Results}
%\vspace{-0.15in}
\end{table}	

To evaluate the efficacy and efficiency of the TinySpeech deep neural networks for limited-vocabulary speech recognition, we leveraged the Google Speech Commands benchmark dataset~\cite{Warden2017}~\footnote{https://research.googleblog.com/2017/08/
launching-speech-commands-dataset.html}, which was specifically for limited-vocabulary speech recognition.  More specifically, the Google Speech Commands benchmark dataset consists of 65,000 one-second verbal utterances of short words as well as background noise.  For comparative purposes, we also evaluated the res15-narrow deep neural network presented in~\cite{Tang2017}, the time delay neural network (TDNN) in~\cite{myer2018efficient}, the trad-fpool13 and tpool2 deep neural networks in~\cite{Sainath2015}, and the PONAS-kws2 deep neural network found using performance-oriented neural architecture search (PONAS)~\cite{anderson2020performanceoriented}, all designed for efficient on-device speech recognition purposes.  The proposed TinySpeech networks were trained using the SGD optimizer in TensorFlow with following hyperparameters: momentum=0.9, learning rate=0.01, number of epochs=50, batch size=64.

The test accuracy, number of parameters, and number of multiply-add operations of the TinySpeech networks in comparison with the other tested networks is shown in Table~\ref{tab_Results}, and a number of interesting observations can be made.  First and foremost, it can be observed that TinySpeech networks achieve significantly lower architectural and computational complexity when compared to the other tested deep neural networks, with TinySpeech-X achieving the highest accuracy and TinySpeech-Z achieving highest architectural and computational efficiency.  More specifically, when compared to the trad-fpool13 network~\cite{Sainath2015}, TinySpeech-X achieved \textbf{4.1\%} higher accuracy while having \textbf{>126.8$\times$} fewer parameters, \textbf{>507.2$\times$} lower weight memory requirements, and requiring \textbf{>11.4$\times$} fewer multiply-add operations.  When compared to the more recent TDNN~\cite{myer2018efficient}, TinySpeech-X achieved \textbf{0.4\%} higher accuracy while having \textbf{>23.2$\times$} fewer parameters, \textbf{>92.8$\times$} lower weight memory requirements, and requiring \textbf{>2.3$\times$} fewer multiply-add operations.  Comparing with res15-narrow~\cite{Tang2017} network, TinySpeech-X achieved higher accuracy (\textbf{0.6\%}) while having \textbf{>3.9$\times$} fewer parameters, \textbf{>15.6$\times$} lower weight memory requirements, and requiring \textbf{>14.6$\times$} fewer multiply-add operations.  Furthermore, when compared to the most recent PONAS-kws2 network, which was found using performance-oriented neural architecture search (PONAS)~\cite{anderson2020performanceoriented}, TinySpeech-X achieved \textbf{0.3\%} higher accuracy while having \textbf{>12.1$\times$} fewer parameters, \textbf{>48.4$\times$} lower weight memory requirements, and requiring \textbf{>15.4$\times$} fewer multiply-add operations.

Let us know explore the performance of TinySpeech-Y and TinySpeech-Z to investigate the trade-off between accuracy and efficiency made during the machine-driven design exploration process.  In the case of TinySpeech-Y, when compared to the trad-fpool13 network~\cite{Sainath2015}, TinySpeech-Y achieved \textbf{3.1\%} higher accuracy while having \textbf{>224.5$\times$} fewer parameters, \textbf{>898$\times$} lower weight memory requirements, and requiring \textbf{>19.2$\times$} fewer multiply-add operations.  Comparing with res15-narrow~\cite{Tang2017} network, TinySpeech-Y had \textbf{>6.9$\times$} fewer parameters, \textbf{>27.6$\times$} lower weight memory requirements, and required \textbf{>24.6$\times$} fewer multiply-add operations while achieving \textbf{0.4\%} lower accuracy.  In the case of TinySpeech-Z, when compared to the trad-fpool13 network~\cite{Sainath2015}, TinySpeech-Z achieved \textbf{1.9\%} higher accuracy while having \textbf{>507$\times$} fewer parameters, \textbf{>2028$\times$} lower weight memory requirements, and requiring \textbf{>48$\times$} fewer multiply-add operations.

We further compare the performance of TinySpeech-Y and TinySpeech-Z with a state-of-the-art efficient deep speech recognition network based on the Legendre Memory Unit (LMU)~\cite{voelker2019lmu}, a new type of recurrent neural network that was demonstrated to achieve orders of magnitude fewer parameters than LSTMs.  In particular, we compare with LMU-4~\cite{blouw2020hardware} since the authors of that study performed substantial efficient optimizations ranging from 4-bit weight quantization to pruning 91\% of the weights, and was able to achieve comparable model size as TinySpeech-Y and comparable accuracy as TinySpeech-Z.  For consistency, we compare TinySpeech-Y and TinySpeech-Z using the same methodology as that proposed by the authors of~\cite{blouw2020hardware} using test accuracy and model size.  It can be seen from Table~\ref{tab_Results2} that TinySpeech-Y achieves \textbf{0.9\%} higher accuracy than LMU-4 at a slightly smaller model size, while TinySpeech-Z is \textbf{>2.2$\times$} smaller than LMU-4 in terms of model size at a \textbf{0.3\%} lower accuracy.  These results illustrate that even when compared to a new type of neural network that has underwent substantial efficiency optimizations, the proposed TinySpeech network architectures leveraging attention condensers can still achieve strong balance between accuracy and efficiency.

\begin{table}[h]
	\centering
	\caption{Test accuracy and model size of TinySpeech-Y and TinySpeech-Z in comparison to LMU-4~\cite{blouw2020hardware}, a state-of-the-art efficient network based on the Legendre Memory Unit (LMU). Best results are in \textbf{bold}.}
	\begin{tabular}{p{3cm}c|cc}
		\hline
		\textbf{Model}& \textbf{Test Accuracy} & \textbf{Model Size (kbits)} \\
		\hline %\hline
		LMU-4~\cite{blouw2020hardware}	&	92.7\%	&	49 \\
		TinySpeech-Y		& 	\textbf{93.6\%}	&	48.8 \\
		TinySpeech-Z		& 	92.4\%	&	\textbf{21.6} \\
		\hline
	\end{tabular}\\
	\label{tab_Results2}
%\vspace{-0.15in}
\end{table}	

Finally, we now explore the performance of TinySpeech-M, which was produced under microcontroller operation constraints imposed during the machine-driven design exploration process.  It can be observed from Table~\ref{tab_Results} that TinySpeech-M achieved \textbf{1.4\%} higher accuracy than the trad-fpool13 network~\cite{Sainath2015} while having \textbf{$\sim$291$\times$} fewer parameters, \textbf{$\sim$1164$\times$} lower weight memory requirements, and requiring \textbf{>28.4$\times$} fewer multiply-add operations.  These experimental results demonstrate the efficacy of leveraging attention condensers and machine-driven design exploration to build highly-efficient deep neural network architectures tailored for on-device speech recognition by striking a strong balance between accuracy, computational complexity, and architectural complexity.

\section{Conclusions}
%\vspace{-0.15in}
\label{Conclusions}
In this study, we introduce the notion of attention condensers for building highly-efficient yet high-performance deep neural networks for on-device speech recognition for edge scenarios.  By jointly modeling local activation relationships and cross-channel activation relationships within a unified condensed embedding, attention condensers can act as self-contained, stand-alone modules that can be leveraged within a deep neural network architecture to reduce the quantity of larger stand-alone convolution modules needed to achieve a high level of accuracy.  We demonstrated the efficacy of the proposed attention condensers by introducing and evaluating TinySpeech, low-precision deep neural networks comprising largely of attention condensers tailored specifically for limited-vocabulary speech recognition.  Experimental results showed that the proposed TinySpeech networks were able to achieve significantly lower architectural and computational complexity when compared to previously proposed deep neural networks designed for limited-vocabulary speech recognition.

Given the promising results associated with the proposed attention condensers, we will explore its efficacy in the future for creating highly-efficient deep neural networks for other tasks such as visual perception, natural language processing, and drug discovery to study its potential for empowering a wide range of TinyML applications. Furthermore, we aim to investigate and explore a variety of different embedding structure designs, condensation designs, and expansion designs, and study their efficacy for further improving the tradeoff between accuracy and efficiency.  Finally, given recent findings that self-attention architectures may exhibit greater robustness to adversarial perturbations~\cite{hsieh-etal-2019-robustness,hendrycks2019natural}, we aim to explore whether leveraging attention condensers can lead to improved adversarial robustness.

\section{Broader Impact}
%\vspace{-0.15in}
\label{BroaderImpact}
TinyML (tiny machine learning) has been heralded by many from academic to industry as the disruptive technology that will empower the widespread adoption and proliferation of machine learning in society as ubiquitous technology.  By bridging the gap between machine intelligence and low-power embedded hardware, TinyML opens the door for numerous applications of on-device machine learning such as smart manufacturing, smart grids, low-cost advanced driver assistance systems and autonomous vehicles, intelligent assistive technologies for the elderly and individuals with impairments, smart cities, intelligent micro-satellites, wearable human-machine interfaces, intelligent supply chain and retail, personal health monitoring, smart agriculture monitoring, just to name a few.  By facilitating for tetherless machine learning on the edge, TinyML can enable real-time decision-making without the need for continuous connectivity to the cloud,  which is critical for empowering a wide range of applications where privacy, security, dependability, cost, and real-time considerations are critical factors to deployment.  Furthermore, by greatly reducing the computational resources needed to operate, TinyML can potentially improve equity and accessibility by enabling machine learning to be leveraged by the masses, be they large corporations with large financial resources or small companies and individuals with limited financial budgets.  As such, the emergence of TinyML can have significant socioeconomical implications given its role as one of the core enablers for ubiquitous machine learning.

Through the exploration of new mechanisms such as attention condensers and investigating their potential for enabling highly-efficient deep neural networks that can operate in an untethered manner on low-power embedded devices, we believe that
the insights gained through such explorations will give us a deeper understanding how such mechanisms behave and can be improved.  The results of such insights contribute to the advancement of TinyML by allowing the community to discover new ways to build more efficient deep neural networks to use in real-world TinyML applications that impact society at large.

\footnotesize
\bibliographystyle{IEEEtran}
\bibliography{tinyspeech}

\end{document}